\documentclass[aps,prl,onecolumn,superscriptaddress,showpacs,amsfonts,amsmath,amssymb]{revtex4-1}\usepackage{natbib}
\usepackage[]{inputenc}
\usepackage[]{graphicx}
\usepackage{bm,color,subfigure,amsmath}
\begin{document}

\title{Dynamical magnetic skyrmions}

\author{Y. Zhou}
\affiliation{Department of Physics, The University of Hong Kong, Hong Kong, P. R. China}
\affiliation{Center of Theoretical and Computational Physics, Univ. of Hong Kong, P. R. China}

\author{E. Iacocca}
\affiliation{Physics Department, University of Gothenburg, 412 96, Gothenburg, Sweden}

\author{A. Awad}
\affiliation{Physics Department, University of Gothenburg, 412 96, Gothenburg, Sweden}

\author{R. K. Dumas}
\affiliation{Physics Department, University of Gothenburg, 412 96, Gothenburg, Sweden}

\author{F. C. Zhang}
\affiliation{Department of Physics, Zhejiang University, Hang Zhou, P. R. China}
\affiliation{Department of Physics, The University of Hong Kong, Hong Kong, P. R. China}

\author{H. B. Braun}
\affiliation{UCD School of Physics, University College Dublin, Dublin 4, Ireland}
\affiliation{Theoretical Physics and Department of Materials, ETH Z{\"u}rich, CH-8093 Z{\"u}rich, Switzerland}

\author{J. \AA{}kerman}
\affiliation{Physics Department, University of Gothenburg, 412 96, Gothenburg, Sweden}
\affiliation{Material Physics, School of ICT, Royal Institute of Technology, Electrum 229, 164 40, Kista, Sweden}

\begin{abstract}
\textbf{Spin transfer torque (STT) affords magnetic nanodevices the potential to act as memory, computing, and microwave elements operating at ultra-low currents and at a low energy cost. Spin transfer torque is not only effective in manipulating well-known magnetic structures, such as domain walls and vortices, but can also nucleate previously unattainable nano-magnetic objects, such as magnetic droplets and skyrmions. While the droplet and the skyrmion are both solitons, the former is inherently dynamic and non-topological, whereas the latter is static but topologically protected. Here we show that it is possible to combine these properties into a novel topologically protected \textit{dynamical skyrmion}, which adds additional degrees of freedom, and functionality, to both droplet and skyrmion based applications. Unlike static skyrmions, the dynamical skyrmion can be nucleated and sustained \textit{without} Dzyaloshinskii–-Moriya interaction (DMI) or dipole-dipole interaction (DDI), and is a generic soliton solution independent of STT and damping once nucleated. In the presence of large DMI, the dynamical skyrmion experiences strong breathing with particular promise for skyrmion-based memory and microwave applications.}
\end{abstract}
\maketitle

There has been a recent rapid increase in the research of magnetic skyrmions \cite{Roessler2006,Heinze2011,Schulz2012,Seki2012,Nagao2013,Ritz2013,Milde2013,Nagaosa2013review,Brataas2014}, which are particle-like topological solitons originally discovered in bulk ferromagnets lacking inversion symmetry, such as the non-centrosymmetric MnSi and FeCoSi~\cite{Muhlbauer2009,Yu2010} and later also in thin films of similar materials~\cite{Yu2010,Yu2011}.  The magnetic skyrmion's spin texture results from a balance between the ordinary ferromagnetic exchange coupling, the Dzyaloshinskii–-Moriya interaction (DMI), and the Zeeman energy from the applied field. Very recently, skyrmions were also proposed as the next generation magnetic information carriers in ultrathin magnetic nanowires where asymmetric interfaces provide the necessary DMI~\cite{Sampaio2013,Fert2013,Iwasaki2013}. As information carriers, it is vital to nucleate isolated skyrmions in opposition to the skyrmion lattice phase observed for non-centrosymmetric thin films~\cite{Yu2011}. Such isolated skyrmions were recently demonstrated experimentally by using spin-polarized tunneling microscopy (STM) at zero field in one monolayer of Fe grown in Ir(111)~\cite{Romming2013}.

In parallel with this rapid development, a novel dynamic, dissipative, and non-topological magnetic soliton, the so-called magnetic droplet~\cite{Hoefer2010}, was very recently experimentally demonstrated~\cite{Mohseni2013} using spin transfer torque (STT) in nano-contact spin torque oscillators (NC-STOs) with perpendicular magnetic anisotropy (PMA) free layers. While originally considered a theoretical curiosity, only stable in magnetic PMA films with zero spin wave damping~\cite{Ivanov1977,Kosevich1990}, the advent of STT~\cite{Slonczewski1996,Slonczewski1996,Berger1996} made it possible to locally create an effectively loss-less spin wave medium~\cite{Rippard2010,Mohseni2011} with the required material properties for droplet nucleation, control, and manipulation~\cite{Hoefer2010,Hoefer2012}.

While droplets and skyrmions have up to this point been studied entirely separately, with little cross-fertilization between the two, they are in fact strongly related. The fundamental properties that so far separate them are their dynamical, topological, and dissipative character. The droplet is dynamic in the sense that all its spins precess at a single frequency; in a skyrmion the spin texture stays static and only its spatial extent can be varied by external factors e.g. ac electromagnetic waves and thermal gradients~\cite{Mochizuki2012,Onose2012,Mochizuki2014}. The droplet is non-topological with a skyrmion number of zero; the skyrmion is topologically protected and has a skyrmion number of 1. 

Despite these seemingly mutually exclusive properties, we here demonstrate how several of these characteristics can be successfully combined, yielding at the same time a dynamical and topologically protected magnetic soliton \textcolor{black}{- a so-called dynamical skyrmion (DS). We develop an analytical theory to demonstrate that the DS is a generic solution, and its sustenance does not depend on DMI, DDI, and current-associated Oersted field. In the presence of DMI, however, the DS shows great potential for both strong microwave signal generation and novel skyrmionic functionality.}

\section*{Dynamical skyrmions as a generic solution}

As our starting point, we micromagnetically~\cite{Vansteenkiste2011} model a NC-STO in zero applied field with an ultra-thin Co layer with strong PMA, without any DMI, and neglecting dipole-dipole interactions (DDI) (for the micromagnetic details, see Methods). As expected, we nucleate an ordinary magnetic droplet above a critical current given by the Slonczewski instability to auto-oscillations \cite{Hoefer2010}. The droplet is characterized by a reversed core, with all spins precessing in phase, and a trivially zero skyrmion number  ${\cal S} = {1 \over 4 \pi} \int \! \int n \, dx\, dy$
($n=  {\bf m}\cdot (\partial_x {\bf m} \times \partial_y {\bf m})$ being the topological density \cite{Moutafis2009,Braun2012} and
${\bf m}$ the magnetization unit vector).

\begin{figure}[t]
\includegraphics[width=7in]{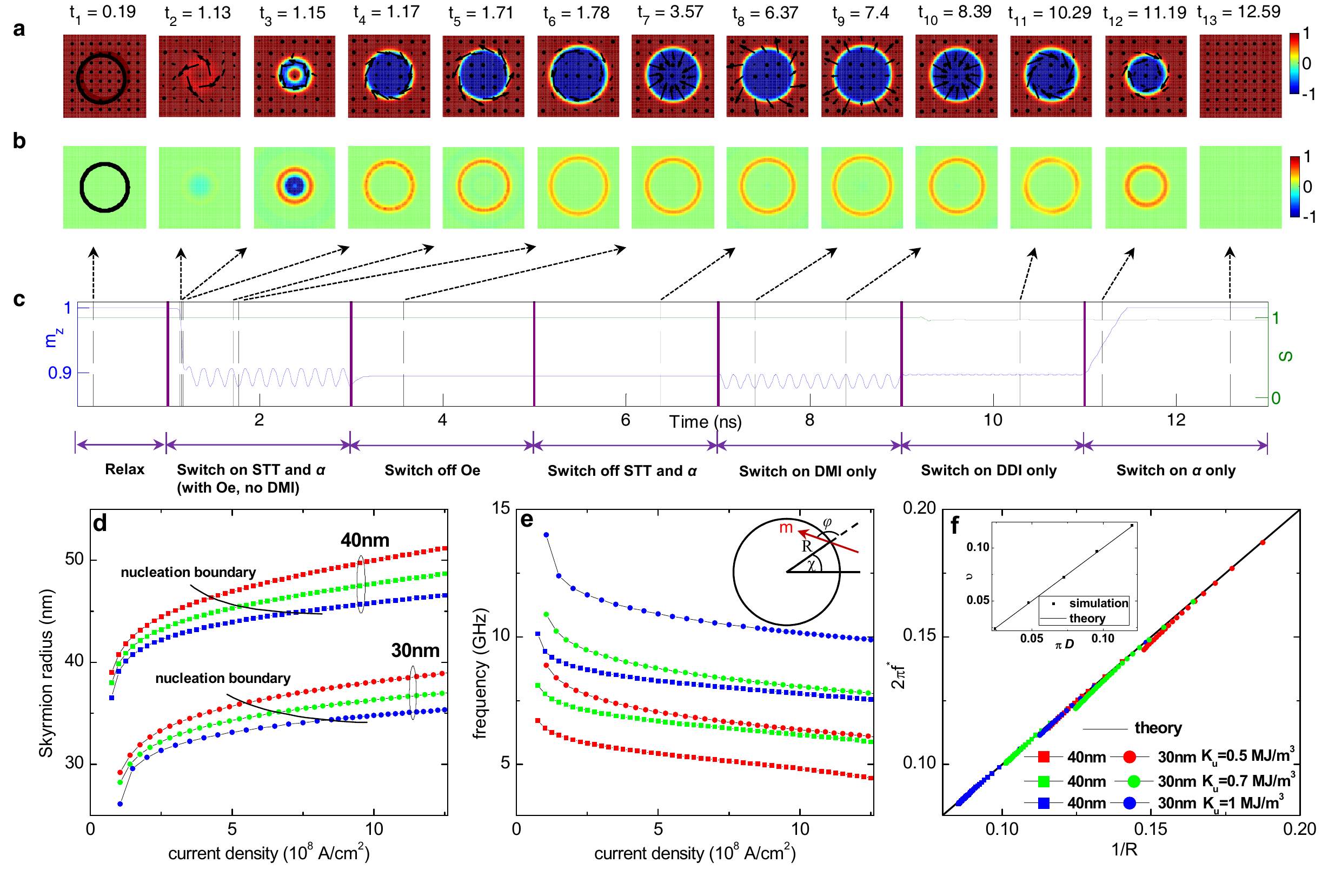}
\caption{\textbf{Nucleation and tuning of a dynamical skyrmion.} A micromagnetic simulation of a NC-STO with radius 40 nm and at $J = 6\times10^8$~A/cm$^2$ showing the nucleation of a dynamical skyrmion (DS) without DMI and DDI  from an initial FM state and the subsequent tuning of the DS by STT, $\alpha$, DMI, and DDI. \textbf{a} top-view of the spin structure at thirteen different times of the simulation; the black circle indicates the NC; \textbf{b} top-view of the topological density at the same times as in \textbf{a}; \textbf{c} time-trace of the out-of-plane $m_z$ component averaged over the simulation area and time-trace of the skyrmion number; dashed vertical lines correspond to the thirteen snapshots above; \textbf{d} shows the DS radius vs current density for two NC sizes and three different PMA strengths; \textbf{e} shows the corresponding DS frequency; Inset shows the coordinate used in the theory. \textbf{f} shows the linear scaling law between the DS frequency and inverse radius, for currents at threshold and below, in excellent agreement with the analytical prediction. The inset shows the DMI-induced breathing amplitude $\nu=(R_{\rm max} - R_{\rm min}) / (R_{\rm max} + R_{\rm min})$ as a function of $D_0$, in which the simulation results are again in excellent agreement with the theory. For panels \textbf{d}-\textbf{f}, $\alpha=0.3$, D=0, with Oersted field.
}
\label{fig:fig1}
\end{figure}

However, if we increase the current density we find that there exists a second, higher, threshold current above which we, instead of a droplet, excite a precessing object with a \emph{non-zero} skyrmion number. Fig.~\ref{fig:fig1} presents the simulation results where panel Fig. 1a displays snap shots of the top view of the free layer magnetization, Fig. 1b the corresponding topological density, and Fig. 1c the time trace of the average out-of-plane magnetization component $m_z$ and the corresponding skyrmion number (see also Supplementary Movie S1). After an initial relaxation, a current density of  $J = 6\times10^8$~A/cm$^2$ is switched on. The system first forms a doughnut structure ($t_2=1.13$ ns and $t_3=1.15$ ns) due to the Oersted (Oe) field and eventually ($t_4=1.17$ ns) forms a dynamically precessing object, characterized by a fully reversed core, spins precessing along its perimeter, and $\mathcal{S}=1$. This new state retains the dynamical precessing character of the droplet, while having the topology of a skyrmion, and can hence be described as a \emph{dynamical skyrmion} (DS). Contrary to the droplet, where all spins precess in phase, the spins along the perimeter of the DS undergo a $2\pi$ rotation, leading to continuous transformations between hedgehog and vortex-like spin textures \cite{Thiaville2012}. The spatially averaged in-plane magnetisation is hence constant in time. However, because of the large static Oe field, the DS experiences a time-varying Zeeman energy and as a consequence exhibits breathing at exactly the precession frequency; if we artificially turn off the Oe field ($t=3$~ns) the breathing disappears. The breathing causes a strong variation in $m_z$ that allows us to unambiguously determine the precession frequency. This behaviour is reminiscent of the breathing observed in quasi-1D magnetic droplet soliton pairs with non-zero chirality~\cite{Iacocca2014} and the DS indeed transforms into such a soliton pair if the lateral dimension of the simulation area is reduced to that of the NC (not shown).

If we simultaneously turn off both the current and the damping ($t=5$~ns) the DS remains stable over time, indicating that it is a novel generic conservative soliton solution for a DMI-free PMA film in zero field. This is hence the topologically nontrivial counterpart of the well known non-topological magnon drop described in the 1970s~\cite{Ivanov1977,Kosevich1990}. If we add a moderate DMI of 0.2 mJ/m$^2$ ($t=7$~ns) $m_z$ again varies in time and breathing resumes. If instead of DMI and Oe field we only turn on DDI ($t=9$~ns), minor breathing is again observed, but this time at twice the precession frequency~\cite{Kosevich1990}. Finally, if only damping is turned on ($t=11$~ns), the DS transforms into a uniform ferromagnetic state, accompanied with an emission of spin waves as shown in the Supplementary Movie S1. This again clearly demonstrates that the DS is not an excited state or an eigenmode of a static skyrmion, but a fundamentally novel solution in the form of a precessing skyrmion \emph{without} the necessity of DMI or DDI.

As seen in Fig.~\ref{fig:fig1}d-e the NC size and the strength of the PMA ($K_u$) controls the overall frequency and radius of the DS, which can then be further actively tuned by the drive current: higher current increases the DS radius and red shifts its frequency. Remarkably, it is also possible to reduce the drive current well below the nucleation current without losing the DS. Similar to the droplet, the sustaining current of the DS can hence be much smaller than its nucleation current, resulting in a very large hysteresis in current. 

These numerical results can be understood in terms of an effective analytical model. The cylindrical symmetry of the DS lends itself to a description  with
the skyrmion radius $R(t)$ and the relative azimuthal angle $\varphi(t)$ as dynamical
variables (for details see inset of Fig. 1e). As described in Supplementary Section S1, the Landau-Lifshitz-Gilbert-Slonczewski equation leads to the following effective equations of motion,
\begin{subequations}
 \label{Rphieqn}
 \begin{align}
 \dot  R  &=  F(\varphi, R)  - \alpha  \dot \varphi   +  \tilde \sigma(I)\tilde \Theta(a_c - R),  \\
  \dot \varphi &=   {1\over  R} + \alpha  \dot R
  \end{align}
\end{subequations}

The dot indicates a derivative with respect to time which is measured in units of the inverse
anisotropy frequency $M_s/2 \gamma K_u$. The skyrmion radius $R$ is defined as the radius of the circle where $m_z=0$ and is measured in units of the domain wall width $\delta_0=\sqrt{A/K_u}$. $F(\varphi, R)$ is a function containing contributions from DMI, Oe-fields, and DDI, $\alpha$ is the damping constant,  $\tilde \sigma(I)$ denotes the dimensionless
spin-torque amplitude, and $\tilde\Theta(x)$ denotes the Heaviside step function smoothed over a length $\delta_0$, and $a_c$ is the radius of the
NC measured in units of $\delta_0$ (for details see Supplementary Section S1).
 Equation (\ref{Rphieqn}) holds under the assumption that  $R > 1$.

The first order differential equations (\ref{Rphieqn}) capture the essence of the skyrmion dynamics observed
in numerical simulations shown in Fig. 1.
In the idealized case of an undamped and undriven system, $\alpha=\tilde \sigma=0$, the
instantaneous skyrmion radius and precession frequency obey the simple relation,
\begin{equation}
2\pi f^* R = 1
\label{omegaR},
\end{equation}
where $f^*=\dot \varphi/ {2\pi}$ is measured in units of the anisotropy frequency.
As illustrated in Fig. 1f, our micromagnetic simulations follow this relation very closely
without any adjustable parameters and
even in presence of nonvanishing currents
and finite damping, if the time average is taken over one period.
Considering the fact that we map a micromagnetic system with
many degrees of freedom onto an effective system with only two dynamical variables, such quantitative
agreement is quite remarkable.

In the absence of DMI, Oe-fields, and DDI,  $m_z$ is conserved, and $F=0$. In this case, both $R$ and $\dot \varphi$ are constant in time, and a skyrmion with constant radius
is stabilized by nonlinear uniform precession around the easy-axis, which satisfies ${\cal S}=1$ at
all times.
It is remarkable that such a precessional DS can exist even in the absence
of DMI, which is also observed in the full micromagnetic simulations ({\it cf.}
Fig. 1a-c in the interval $5\,{\rm ns} \leq  t <  7\,{\rm ns}$).  Such behaviour
persists in the presence of damping and compensating spin torque
as is seen from
the interval $3\,{\rm ns} \leq  t < 5\,{\rm ns}$. In this case, equation (\ref{Rphieqn}) can be reduced to
$(1+\alpha^2) \dot R = - \alpha/R + \tilde \sigma \tilde \Theta$ with solutions asymptotically converging to a time independent radius  $R \approx \tilde a_c + \delta_0$. For $\tilde \sigma =0$, the dynamics yields
an ever shrinking skyrmion which eventually disappears due to lattice discreteness effects. This
latter effect is seen in the simulations after $t= 11$ ns.

If in turn one of these interactions is nonzero, then $F\neq 0$, $m_z$ is no longer
conserved, and the time dependence of the precession frequency
is obtained from
$\ddot \varphi + F(\varphi) \dot \varphi^2=0$ (Supplementary Section S1). In the case of  vanishing Oe-fields, this is
$R$-independent and can
readily be integrated.
The solutions exhibit nonuniform precession
and by virtue of equation (\ref{omegaR}), breathing of the skyrmion.
Such breathing is indeed seen in the interval $7\,{\rm ns} \leq  t < 11\,{\rm ns}$
of Fig. 1a-c (also refer to Supplementary Section S1).

A breathing skyrmion oscillates between radii $R_{\rm max}$ and $R_{\rm min}$ with corresponding
relative azimuthal angles $\varphi_{\rm max}$ and $\varphi_{\rm min}$, respectively,
which are in turn determined by the zeroes of $F(\varphi)$.
If only DMI contributes
then the reversal occurs at $\varphi_{\rm max}=0$ and
 $\varphi_{\rm min}=\pi$ ($t_9$, $t_{10}$ in Fig. 1). Thus at the largest extension, the skyrmion is in a hedgehog configuration
favoured by DMI. For Oe-field $h_{\rm Oe} < 0$ ({\it i.e.} current flowing downwards as in the simulations)
we have $\varphi_{\rm max}=\pi/2$ and $\varphi_{\rm min}=3 \pi/2$,
and thus at the largest radius, the skyrmion is in the energetically favoured  tangential configuration oriented in a right handed fashion around the current ($t_5$, $t_6$ in Fig. 1). In these cases, breathing and precession frequency coincide. Finally, if only DDI contributes, then both time reversal and parity are unbroken and a skyrmion with maximal radius occurs at two values  $\varphi_{\rm max}=\pi/2,  3\pi/2$, while $\varphi_{\rm min}=0, \pi$. As a consequence, the radial oscillation frequency is twice the azimuthal frequency, which is indeed observed in the simulations in the interval $9\,{\rm ns} <  t \leq 11\,{\rm ns}$.

In the case of DMI only, we also can easily compute the breathing amplitude $\nu$ upon time integration of equation (\ref{Rphieqn}a) over a half period. This yields $\nu=(R_{\rm max} - R_{\rm min}) / (R_{\rm max} + R_{\rm min})=  \pi D_0$
 with $D_0 = D/{\cal E}_0$ with ${\cal E}_0$ the domain wall energy per unit area. This relation is followed strikingly well in the presence of both current and damping as shown in the inset of Fig. 1f.

Interestingly, the critical current density for DS nucleation can be reduced dramatically (more than one order of magnitude)  by using a smaller damping $\alpha$, smaller $K_u$, and larger NC size (See Supplementary Section S2). Under these conditions, it becomes possible to create DS with relatively small current density in experiments. Furthermore, we confirmed that the above results and conclusions using a damping of $\alpha$=0.3 remain qualitatively the same if a much smaller damping of $\alpha$=0.05 is used (see Supplementary Section S2).

\section*{Dynamical skyrmions in the presence of large DMI}

\begin{figure}[t]
\includegraphics[width=7in]{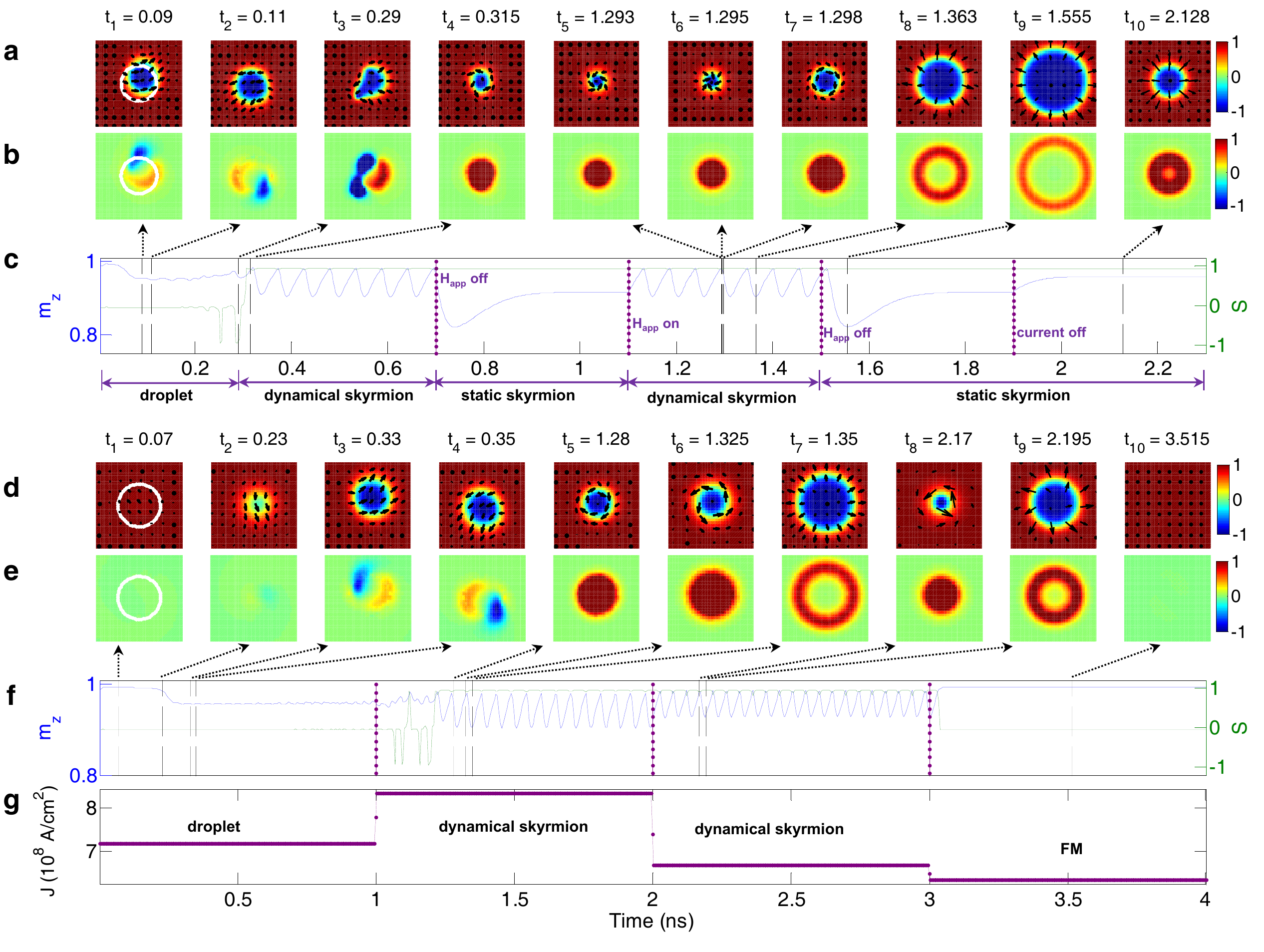}
\caption{\textbf{Nucleation, field- and current-toggling of a DS in presence of DMI.} Panels \textbf{a-c} show, respectively: the top-view of the spin structure at selected simulation times where the white circle indicates the 15nm NC; the topological density at the same times where the color-scale is normalized; the time-trace of the out-of-plane magnetization component $m_z$ averaged over the simulation area; and time-trace of the skyrmion number (green). Panels \textbf{a-c} show droplet nucleation at $J = 8.3\times10^8$~A/cm$^2$ and $\mu_0H_a$ = 0.3 T, which remains stable for several periods until about $t$ = 0.2 ns when it becomes increasingly susceptible to anti-skyrmion perturbation ($\mathcal{S}<0$). These perturbations eventually ($t$ = 0.3 ns) give way to the formation of a DS with $\mathcal{S}=1$. When the applied field is turned off at $t$ = 0.7 ns, the DS rapidly dissipatively shrinks into a static skyrmion. If the field is again turned on, the skyrmion can be transformed into a DS in a reversible manner. Finally, if both the field and the current are turned off, the static skyrmion contracts to its equilibrium size given by the material parameters. Panels \textbf{d-f} show a similar DS nucleation as a function of current and fixed field. The current density is varied as follows: $J$ = 7.2$\times10^8$ A/cm$^2$ for 0 $<$ t $<$ 1 ns, 8.3$\times10^8$ A/cm$^2$ for 1 ns $<$ t $<$ 2 ns
, 6.7$\times10^8$ A/cm$^2$ for 2 ns $<$ t $<$ 3 ns, and 6.3$\times10^8$ A/cm$^2$ for 3 ns $<$ t $<$ 4 ns. Panel \textbf{g} schematically shows the current pulses applied during the simulation. See Supplementary Movie S2 and S3 for the entire process.}
\label{fig:fig2}
\end{figure}

We now turn to the interesting case when the DS is nucleated and sustained in a material with large DMI. This situation is of particular applied importance as such systems are known to support static skyrmions and should hence allow for the interaction of droplets, skyrmions, and dynamical skyrmions. Figure~\ref{fig:fig2}a-d shows the rapid nucleation of a magnetic droplet soliton exhibiting its typical characteristics of precessing spins along its perimeter. Because of the large DMI, the spin structure is substantially perturbed ($t_1$ and $t_2$) compared to the situation where DMI is absent. At times $t_1$ and $t_2$, regions of weak non-zero topological density are found to rotate around the droplet perimeter. While the droplet is stable for a number of periods the topological perturbations continue to grow in amplitude and almost drive the formation of an anti-skyrmion ($\mathcal{S}\rightarrow-1$ at $t_3$) just before the system switches into a stable $\mathcal{S}=1$ state at $t_4$. The transition mechanism occurs through a Bloch point pair or a single Bloch point on the surface (See Supplementary Section S3 for details).

If the applied field along +$z$ is turned off, the DS relaxes into a static hedgehog skyrmion; if the applied field is again turned on, the DS reforms as the precession restarts. The DS to static skyrmion transition is hence entirely reversible, which is a natural consequence of their identical topology. Finally, if both field and current are turned off, a smaller static hedgehog skyrmion remains with its size given by the DMI and material parameters of the simulation. From the STT provided by a non-zero current density, one can controllably tune the size of the static skyrmion, where a positive current density increases its size and a negative current density decreases it.

\begin{figure}[t]
\includegraphics[width=5.5in,trim= 0 0 00 0, clip=true]{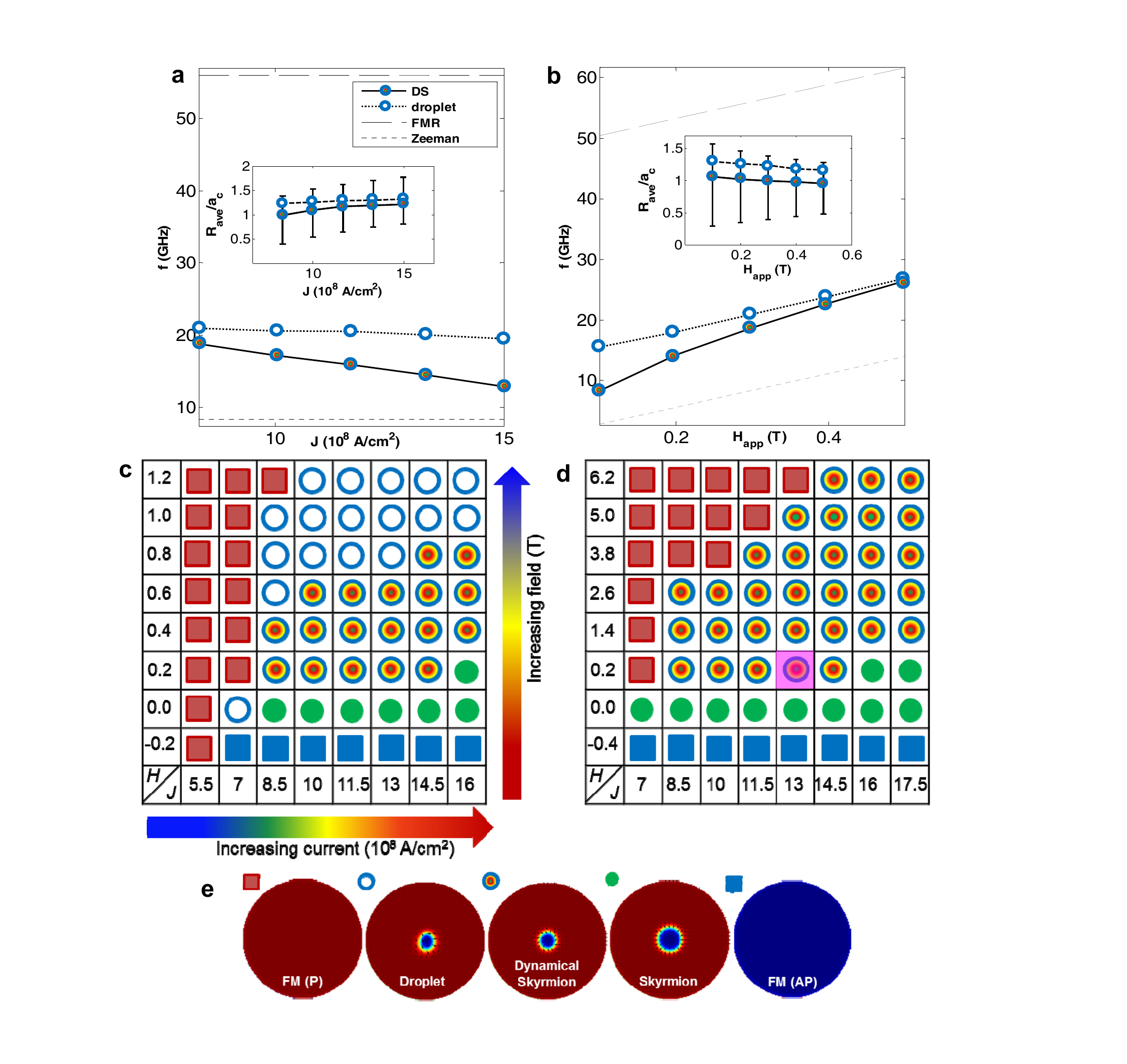}
\caption{\textbf{Frequency and stability of DS.} The frequency of the dynamical skyrmion is shown as solid lines and filled circles for: \textbf{a} $\mu_0H_{a}=0.3$~T while the current density is varied; \textbf{b} $J=8.3\times10^8$~A/cm$^2$ while the applied field is varied. The corresponding droplet frequency (DMI=0) is shown as dotted lines and hollow circles. The Zeeman and FMR frequencies are shown as dashed lines. The insets show the radius of the droplet (hollow circles) and the time averaged radius of the dynamical skyrmion (filled circles) measured in units of NC radius where the error bars indicate the total range of radii values. The DS frequency decreases rapidly with increasing radius (increasing current) in \textbf{a}. As the field increases in \textbf{b}, the dynamical skyrmion becomes stiffer, reducing the breathing and making the dynamics resemble that of the droplet. \textbf{c} Nucleation of a droplet (hollow circle), dynamical skyrmion (filled rainbow circle), and static skyrmion (green filled circle) at different fields and currents. \textbf{d} Stability of the dynamical skyrmion over a very wide range of current and field. Note that the field axis is nonlinear to reach the final collapse of the dynamical skyrmion at very high fields. The DS was nucleated using the conditions highlighted by the pink square.}
\label{fig:phasediagram}
\end{figure}

The DS can also be nucleated by controlling only the drive current density. Figure~\ref{fig:fig2}e-i shows a micromagnetic simulation of the same NC-STO in a constant applied field of 0.3 Tesla, which favours a uniform ferromagnetic state. After a period of weak FMR-like precession, a DMI-perturbed droplet forms and now remains stable for as long as the current density is limited to $J=7.2\times10^8$~A/cm$^2$. When the current density is increased to $J = 8.3\times10^8$~A/cm$^2$ the STT provides enough energy to induce strong topological fluctuations between negative and positive skyrmion numbers to finally switch the system into a stable dynamical skyrmion state. Once the dynamical skyrmion has formed, the current density can be reduced substantially while still sustaining the precession, until a minimum sustaining current density is reached below which the dynamical skyrmion rapidly collapses into a uniform $\mathcal{S}=0$ ferromagnetic state, in a similar fashion as ordinary droplets. As a consequence it is perfectly possible to repeatedly access the uniform, droplet, and dynamical skyrmion states by only controlling the current density. By controlling \emph{both} the drive current density and the applied field, transitions between all four states, including the static skyrmion, can be controlled at will, with the only limitation being the direct transformation of a skyrmion or dynamical skyrmion into a droplet, which requires an intermediate step of a uniformly magnetized state.

In Fig.~\ref{fig:phasediagram}a-b we compare the field- and current-dependent tunability of the DS and the corresponding droplet (the latter simulated by removing the DMI term but otherwise sharing identical conditions). The most salient feature of the DS is a much stronger frequency tunability than the droplet. Whereas the droplet frequency is essentially independent of current density and linearly dependent on the field, the frequency of the DS decreases rapidly and linearly with increasing current density and exhibits a non-linear field dependence, in particular at low fields. Additionally, the DS maximum frequency is bounded by its droplet counterpart.

The key to understanding the much stronger field- and current dependencies lies in the very strong breathing of the DS, which dominates the dynamics at large DMI. While the ordinary droplet is always slightly larger than the NC, and does not vary in size significantly with either current or field, the radius of the dynamical skyrmion can at low fields have a minimum that is less than a third of the NC radius and a maximum that is more than 50\% greater than the NC, as shown by the error bars in the insets in Fig.~\ref{fig:phasediagram}. In other words, the breathing can make the dynamical skyrmion radius vary by more than five times of its minimum size. The very strong breathing will increase the dissipation and the periodic translation of the domain wall making up the dynamical skyrmion perimeter will slow down the overall precession. When the current density is increased in Fig.~\ref{fig:phasediagram}a the maximum radius also increases, further slowing down the precession. However, when the field is increased in Fig.~\ref{fig:phasediagram}b the dynamical skyrmion stiffens, the amplitude of the breathing decreases, and as a consequence, both the the maximum radius and the frequency of the dynamical skyrmion approach those of the droplet.

As observed in Fig.~\ref{fig:fig2} above, the DS exhibits a similar degree of hysteresis as the original droplet, \emph{i.e.} its sustaining current can be much lower than the current needed for nucleation. This hysteresis ensures a minimum degree of stability, which should make the DS sufficiently robust for applications. In Fig.~\ref{fig:phasediagram}, we investigate this stability in more detail and present a nucleation phase diagram in Fig. 3c and a stability phase diagram in Fig. 3d. The nucleation phase diagram presents the final steady state of the simulated system, when both current and field are turned on at $t_0$ and held constant until steady state. Five different end states can be identified: a droplet, a DS, a static skyrmion, and the two trivially saturated states. The stability phase diagram, on the other hand, presents the final steady state at all field and current values after a DS has first been nucleated at the conditions shown in pink. Here, only four different end states are possible as the DS never transforms back directly into a droplet. It is noteworthy that the DS is stable over a very large current and field range, more so than the droplet, which is consistent with its topological protection affording it additional stability.

In summary, we have demonstrated how droplets and skyrmions can be combined to form a dynamical skyrmion, a hitherto unknown topological and dissipative magnetic soliton with great potential for both new physics and direct applications in skyrmionics and NC- based microwave signal generators.

\section{Methods}

\subsection{Micromagnetic simulations}

Micromagnetic simulations are performed for the free layer with the graphics-processing-unit-based tool Mumax3~\cite{Vansteenkiste2011}. The time-dependent spin dynamics follow the Landau-Lifshitz-Gilbert-Slonczewski equation of motion,
\begin{equation}
{ \label{eq:LLGS}
\frac{d\textbf{m}}{dt}=-|\gamma|\textbf{m}\times{\mu_0{\textbf{H}}_{eff}}+\alpha\textbf{m}\times\frac{d\textbf{m}}{dt}
+|\gamma|\mu_0 M_s\sigma(I)f(\vec{r})\epsilon\textbf{m}\times(\textbf{m}\times\hat{\textbf{z}}), }
\end{equation}
where $|\gamma|/2\pi$=28 GHz/T is the gyromagnetic ratio, $\alpha$ is the Gilbert damping parameter, $\mu_0$ is the magnetic vacuum permeability, $M_s$ is the free layer saturation magnetisation, and $\bf{m}$ is the free layer normalized magnetisation vector. The fixed layer magnetisation vector is along $+\hat{\textbf{z}}$.  $\sigma(I)={\hbar I P\lambda}/{\mu_0M_s^2eV(\lambda+1)}$ is the dimensionless spin torque coefficient where $\hbar$ is the reduced Planck constant, \textit{I} is the electric current, \textit{P} is spin polarization,  $e$ is the elementary charge ($e>0$), $V$ is the free layer volume. $\epsilon^{-1}={1+\dfrac{\lambda-1}{\lambda+1}\hat{m}\cdot\hat{\textbf{z}}}$, where $\lambda$ is the magnetoresistance (MR) asymmetry parameter describing the deviation from sinusoidal angular dependence. $f(\vec{r})$ is the Heaviside step function approximately describing the current distribution, \textit{i.e.}  $f(\vec{r})$=1 for  $r\le a_c\delta_0$  and 0 otherwise. The effective field  $\vec{H}_{eff}$ includes the contribution from Heisenberg exchange, PMA, DMI, and an applied external field along the $+\hat{\textbf{z}}$ direction. The DMI in Mumax3 is assumed to be purely of interfacial origin, giving rise to the energy~\cite{Bogdanov2001}.
\begin{equation}
{ \label{eq:DMI}
\epsilon_{DM}=D\left(m_z\frac{\partial m_x}{\partial x}-m_x\frac{\partial m_z}{\partial x}+m_z\frac{\partial m_y}{\partial y}-m_y\frac{\partial m_z}{\partial y}\right) }
\end{equation}
where $D$ is the DMI constant in units of J/m$^2$.

A NC spin torque oscillator geometry is modeled, based on a pseudo spin valve with PMA fixed and free layers and a current-confining NC of radius $15$~nm placed at the centre of the free layer (except in Fig. 1 where the NC radius is varied between 30~nm and 40~nm). The free layer is assumed to be a $1$~nm thick Co film with radius of 100 nm on a substrate inducing DMI, which can be controlled by varying the layer thickness, and material parameters similar to Ref.~\cite{Sampaio2013}, namely: exchange constant $A = 15$~pJ/m, Gilbert damping $\alpha = 0.3$, spin polarization ratio $P$=0.3, saturation magnetisation $M_s = 580$kA/m, $D = 3$~mJ/m$^2$ and $K_u = 0.7$~MJ/m$^3$ unless otherwise specified. These parameters lead to a domain-wall width of $\delta_0=5.6$~nm. The fixed layer is assumed to be magnetised along the +z direction, inducing a STT with $\lambda=1$ for simplicity. All of our simulations are performed with unit cell size between $1$ and $2$~nm, which is well below $\delta_0$ and the exchange length, ensuring numerical accuracy. For finite temperature simulations in Supplementary Section S4, a fixed time step of $0.1$~fs is used. The simulations are performed with open boundary conditions, \emph{i.e.} damping is set
to be uniform across the whole disk including the boundary (for details see Supplementary Section S5).

\section*{Acknowledgements} Y.Z. thanks the support by the UGC Grant AoE/P-04/08 of Hong Kong SAR government. H.B.B. acknowledges financial support from Science Foundation Ireland (Grant No.  11/PI/1048). This work was partially supported by the ERC Starting Grant 307144 \textquotedblleft MUSTANG\textquotedblright, the Swedish Foundation for Strategic Research (SSF) program "Future Research Leaders", the Swedish Research Council (VR), and the Knut and Alice Wallenberg Foundation. Johan \AA kerman is a Royal Swedish Academy of Sciences Research Fellow supported by a grant from the Knut and Alice Wallenberg Foundation.

\end{document}